# Analysis tools for single-monomer measurements of self-assembly processes

Maria Hoyer[1], Alvaro H. Crevenna[1,2], Radoslaw Kitel[3,4], Kherim Willems[5], Miroslawa Czub[3], Grzegorz Dubin[4], Pol Van Dorpe[5], Tad A. Holak[3] and Don C. Lamb[1]*

[1] Department of Chemistry, Center for NanoScience, Nanosystems Initiative Munich (NIM) and Center for Integrated Protein Science Munich (CiPSM), Ludwig-Maximilians University Munich, Munich, Germany.

[2] Epigenetics and Neurobiology Unit, EMBL Rome, Monterotondo, Italy

[3] Department of Organic Chemistry, Faculty of Chemistry, Jagiellonian University, Gronostajowa 2, 30-387 Krakow, Poland

[4] Malopolska Centre of Biotechnology, Jagiellonian University, Gronostajowa 7a, 30-387 Krakow, Poland

[5] imec, Kapeldreef 75, B-3001 Leuven, Belgium

* For correspondence: d.lamb@lmu.de

## Abstract
Protein assembly plays an important role throughout all phyla of life, both physiologically and pathologically. In particular, aggregation and polymerization of proteins are key-strategies that regulate cellular function. In recent years, methods to experimentally study the assembly process on a single-molecule level have been developed. This progress concomitantly has triggered the question of how to analyze this type of single-filament data adequately and what experimental conditions are necessary to allow a meaningful interpretation of the analysis. Here, we developed two analysis methods for single-filament data: the visitation analysis and the average-rate analysis. We benchmarked and compared both approaches with the classic dwell-time-analysis frequently used to study microscopic association and dissociation rates. In particular, we tested the limitations of each analysis method along the lines of the signal-to-noise ratio, the sampling rate, and the labeling efficiency and bleaching rate of the fluorescent dyes used in single-molecule fluorescence experiments. Finally, we applied our newly developed methods to study the monomer assembly of actin at the single-molecule-level in the presence of the class II nucleator Cappuccino and the WH2 repeats of Spire. For Cappuccino, our data indicated fast elongation circumventing a nucleation phase whereas, for spire, we found that the four WH2 motifs are not sufficient to promote *de novo* nucleation of actin.

# Introduction

Protein polymerization plays an important role in both pathological and physiological processes. For example, polymerizing proteins, such as actin, build up the cytoskeleton and are of vital importance for the cell. At the same time, the aggregation of misfolded proteins is associated with Alzheimer's disease or type II diabetes mellitus (*1*). Despite the recent advances in our understanding of elongation processes, many of the details during the early nucleation phases of polymerization are still unknown. As a growing number of proteins are shown to polymerize (*2, 3*), tools to elucidate the mechanism of nucleation become more relevant. In general, nucleation mechanisms are divided into primary and secondary nucleation processes. Primary nucleation is initiated by isolated monomers whereas secondary nucleation requires an initial filament. Most primary nucleation processes studied so far (*4*) have been classified to occur either through a nucleation-elongation mechanism or through a conversion mechanism (Figure 1A). For example, actin polymerization has served as a canonical representative of nucleation-elongation (*5*), although this may depend on the polymerization conditions (*6*), whereas amyloid formation has been reported to follow conversion models (*7-9*). Therefore, analysis tools that elucidate the nucleation mechanism are of wide interest.

Of the various methods available to investigate nucleation, single-molecule techniques are a promising route since they can provide direct, real-time measurements of single monomer association and dissociation events occurring during the growth process (*10*). In particular, single-molecule fluorescence techniques combined with the use of zero-mode waveguides (ZMW) (*11*) have opened the possibility to observe the early stages in oligomer assembly directly. Similarly, single events such as binding and dissociation of monomers on individual oligomers can be measured via scattering (*12*). By using single-molecule techniques, binding events can be visualized as step-wise signal increases over time and dissociation events show a step-wise, decrease in signal (Figure 1B). In an often used approach, the residence-time at each monomer number is pooled together and typically than fit to an exponential function to determine the lifetime of the oligomer at a particular monomer number. This is repeated for all monomer numbers. This method is often referred to as a dwell-time-analysis and has been applied in many biophysical studies (*13, 14*). The dwell-time-analysis provides the microscopic rate for each step and can, therefore, reveal the unique underlying mechanism. However, the accuracy of the analysis, and thereby the identification of the nucleation mechanism, depends on the quality of the data and sufficient sampling of the different steps. For example, to estimate the rate of a single transition accurately, a sampling rate has to be chosen that is at least ten times as fast as the rate to be estimated. For some techniques such as optical tweezers, the sampling frequency is not a problem as it is only dictated by the readout speed of a position-sensitive detector. However, getting sufficient statistics is more difficult due to the low throughput of these types of measurements. For fluorescent samples, high sampling frequency comes at the expense of increased noise and/or an increased probability of photobleaching, which complicates the analysis and shortens the total experimental time.

Besides the wealth of data provided by single-molecule techniques, information on the underlying mechanism of nucleation and growth can be extracted equally from the average time-course of filaments growing on many surface-bound nucleator proteins, for example, by using total-internal reflection microscopy (TIRFM) (*15*), circular dichroism (*16*), or other spectroscopy methods (*17-19*). By only measuring filaments growing via the nucleator protein of interest tethered to the surface, spontaneous oligomerization events occurring in solution can be excluded. This approach provides the possibility to synchronize the starting point of the oligomerization process.

In this work, we developed and tested analysis methods for kinetic data of filament growth on the single oligomer level as well as for averaged but synchronized filament growth. Our objective is to i) recover the correct on- and off-rates of the single monomer addition and dissociation events, and ii) recover the correct assembly mechanism via the relative differences between monomer binding events at different oligomer sizes. Thereby, we wish to distinguish between the two main mechanisms: a nucleation mechanism with a slow on-rate or high off-rate in the beginning until the polymerization phase is reached, and a conversion mechanism where a single slow on-rate or high off-rate defines a bottleneck for successful polymerization. We test our developed methods by simulating the growth processes of single oligomers with defined binding and dissociation rates of the monomers, dependent on the oligomer size at the time of the binding or dissociation event.

To test the robustness of the analysis tools, we investigated the influence of the SNR, the simulated measurement rate and the relative difference between the nucleation and polymerization kinetics. Furthermore, since measurements in ZMW and TIRFM depend on dye-labeled monomers, we looked into the distribution of labels per monomer (i.e., the labeling efficiency), as well as the photobleaching of fluorescent labels. Finally, we applied the analysis tools on experimental data of actin nucleation using the formin-homology domain 2 (FH2) of Cappuccino and the WH2 domains of spire (spire-ABCD) as nucleators and compared the results to simulations of unhindered and unsuccessful oligomer growth. For Cappuccino, the data indicated a circumvention of the nucleation phase resulting in unhindered growth as expected. For spire-ABCD, we found that the WH2 domains are not sufficient to promote actin nucleation starting from the purely monomeric species.

## Theory and Simulations

For the development and testing of different analysis methods, we simulated the assembly of single monomers into filaments by sequential monomer addition based on the basic principles of filament formation (*20*). The assembly process is ruled by the kinetics of individual binding and dissociation events. We looked into the two major mechanistic possibilities that have been found to occur during primary nucleation (*21*). First, we looked into a one-step nucleation mechanism, i.e., the formation of a nucleus of defined size (n monomers) that represents the smallest stable structure and allows subsequent polymerization. Hereby, every monomer binding event until the formation of the nucleus is defined by the nucleation kinetics, and the faster polymerization kinetics take effect after the nucleus size has been reached. To simulate slower nucleation kinetics, either the dissociation rate constants of the single monomers can be enhanced, or the association rate constants can be slowed down. It is typically assumed that the association rate constants do not change during a nucleation process (*22*). However, we also looked into the effect of association rates as well to include all possibilities. The second possible mechanism that we investigated is that of a conversion step, where a rearrangement of the oligomer occurs leading to different kinetics. Thus, we tested the different analysis methods for their ability to 1) detect a transition from nucleation-governed kinetics to faster polymerization and 2) to detect a single conversion step and thereby distinguish between the two mechanisms.

## Simulations

We performed stochastic simulations of an assembly process with individual on- and off-rates for each monomer step. The time spent at the current oligomer size as well as whether the next step was an association or dissociation event was randomly selected from an exponential distribution based on the on- or off-rates for the respective oligomer size. This process was continued for each oligomer until the total preselected simulation time was exceeded. Based on the association and dissociation events, a monomer number versus time trace was built using a sampling rate of 100 $s^{-1}$, unless stated otherwise. The selected sampling rate roughly corresponds to the measurement rates of ~10 ms/frame currently achievable with modern cameras. For the nucleation mechanism, the kinetics before reaching the nucleus size were defined by the monomer association rate $k^+_{nuc}$ and the dissociation rate $k^-_{nuc}$, which were treated as identical for oligomers smaller than the nucleus size. When the monomer number reached the nucleus size, the kinetics changed to the polymerization rates $k^+_{poly}$ and $k^-_{poly}$. The second mechanism, a conversion step, was simulated using a single association rate ($k^+_{poly}$) and a single dissociation rate ($k^-_{poly}$) for all monomers with the exception of the conversion step, which shows a slower kinetics $k^+_{conv}$ and $k^-_{conv}$. If not stated otherwise, $k^+_{poly}$ was set to 1 $s^{-1}$ and $k^-_{poly}$ to 0.1 $s^{-1}$, which corresponds to a factor of 100 fold and 1000 fold the sampling time respectively. These rates are in the range of the known rates of pointed end actin polymerization (*23*). Looking at the relative difference between the expected rates and the sampling rate, our results can be used to estimate the necessary sampling rate for an experiment of this kind. In addition, the results can be further evaluated with respect to the sample rate to determine whether the extracted rates are trustworthy. For $k^+_{nuc}$ and $k^+_{conv}$, we chose values that corresponded to 10%, 50%, or 80% of $k^+_{poly}$ to test the sensitivity of the analysis methods for detecting slight changes in the kinetics. Accordingly, we chose $k^-_{nuc}$ and $k^-_{conv}$ to be 2-, 5-, or 10-fold faster than $k^-_{poly}$.

The results of the simulated assembly process were transformed into fluorescence traces by overlaying them with Gaussian noise using a signal-to-noise ratio (SNR) of 2, unless stated otherwise (Figure 1A). We analyzed both single filament traces as well as an average of 1000 traces from individual simulations (Figure 1C). For the single filament traces, we extracted the underlying monomer versus time trace via a step-finding algorithm (see materials and methods for details) (*6*). For slow sampling rates, two fast consecutive association steps could appear as one step with a double step size. Therefore, we used the mode of the step sizes to identify double and triple steps and correct the monomer number accordingly.

## Visitation Analysis and Average rates

As an alternative to the dwell-time analysis, we present two novel analysis tools for investigating the microscopic mechanism underlying filament formation: the visitation analysis and the average-rate analysis. The visitation analysis samples the time the oligomers spend at each oligomer size (Figure 1B, C). The resulting histogram contains information regarding the different rates involved in the single steps of the nucleation process. Steps that are faster or slower than that of their neighboring oligomers are directly visible (Figure 1C). For a polymer that forms via a nucleation-elongation mechanism, the most often detected oligomer sizes are determined by the nucleation phase where oligomers fluctuate between sizes equal to or below the nucleation size. The initial oligomers form and immediately disassemble due to the high thermodynamic barriers that need to be overcome. Upon successful formation of the nucleus, the kinetics changes and elongation occurs. As $k^+_{poly}$ is typically $\gg k^-_{poly}$, the oligomers do not spend much time at specific monomer sizes just above the nucleation size as $k^+_{poly}$ dominates and the oligomers grow (Figure 1B blue line). For a filament that polymerizes through a conversion mechanism, a slow conversion step at a particular monomer number size is required for elongation. Hence, the oligomers spend a significant fraction of the time at this monomer size in comparison to the fast fluctuations at smaller oligomer sizes and the elongation upon the conversion step (Figure 1C orange line).

The second approach is the average-rate analysis. Here, the rates for each monomer addition are calculated from the time it takes for the average trace of 100s to 1000s of growing oligomers to reach the next monomer number (Figure 1D). These average rates are then plotted as a function of oligomer size (Figure 1E). For many oligomers that started growing simultaneously, a nucleation mechanism is directly visible in the shape of the averaged intensity curve since the slope of the curve is less steep in the beginning of the assembly process due to the slower nucleation kinetics (Figure 1D). Since the sum follows the same behavior as the average of many traces, this approach can also be applied to experiments that are synchronized but where single-filament data is not available. Averaging is a useful tool to circumvent the limitations of a low SNR (Figure S2). When single filament data is available, the average intensity from many single-filament traces can still be calculated and both analyses performed. Experimentally, the average signal for the addition of a monomer needs to be known to convert the intensity information into a monomer number. For a growth process that does not reach high monomer numbers on average, fractional average rates in 0.1 monomer steps can also be extracted.

Similar to the visitation analysis (Figure 1C), the average rate (Figure 1E) also reveals signatures that can be associated with either a nucleation-elongation (Figure 1E blue points) or a conversion mechanism (Figure 1E orange points). Nucleation-elongation starts from a low average rate and monotonically increases until it reaches the average elongation rate reflecting the transition from the slow nucleation phase to the faster elongation regime. The average rate from a conversion mechanism first drops around the oligomer size where the conversion occurs and then increases until it also reaches the average elongation rate.

As a frequently employed standard analysis tool, we also tested the dwell-time analysis for its functionality on single-oligomer data. The dwell-time analysis generates a histogram of waiting times until the next monomer binding event occurs for oligomers of different sizes (Figure S1A, B). An exponential fit of the distribution yields the rate of the respective monomer binding event. The on- and off-rates were extracted from an exponential fit of the dwell-time distribution (Figure S1) and plotted against the monomer number to visualize the transition from nucleation to polymerization that occurs once the nucleus size is reached.

### Distinguishing between Nucleation and Conversion Mechanisms

To determine how well the different analysis approaches can distinguish the mechanism of nucleation, data were simulated for both a nucleation mechanism and a conversion mechanism occurring at an oligomer size of 2, 3, 4 and 5 monomers. The visitation analysis and the average-rate analysis are both able to identify the nucleus size and a conversion step accurately, independent of whether the association or the dissociation rate is changed (Figure 2). The dwell-time analysis can identify a change in the association rate for both the nucleation and conversion mechanism (Figure 2F, Figure S1: Dwell-time analysis on an assembly process with a conversion step at 2, 3, 4 or 5 monomers. (A, B): the dwell-time distributions for the first association step (A) and the first dissociation step (B) are shown. The dwell-time distribution for each individual step was fitted with a single-exponential distribution. (C, E): The extracted association rates for a nucleation (C) and a conversion mechanism (E) with different nucleation or conversion sizes. The dwell-time analysis is able to identify the association rates for the individual steps. The association rates during nucleation $k^+_{nuc}$ were set to 50% of $k^+_{poly}$ with no change in the off-rates. (D, F): The extracted dissociation rates for a nucleation (C) and a conversion mechanism (E) with different nucleation or conversion sizes. The dwell-time analysis could not identify the dissociation rate changes during the nucleation or conversion steps. The dissociation rates during nucleation $k^-_{nuc}$ were chosen to be 5x of $k^-_{poly}$ with no change in the on-rates. Error bars for the rates represent the 95% confidence intervals of the exponential fits. Applied to a growth process with a single conversion size, the dwell-time analysis could only identify single steps with a large change in the on-rate (Figure S1D). For both nucleation and conversion, the dwell-time analysis is not sensitive towards a change in the dissociation rates (Figure 2C, Figure S1: Dwell-time analysis on an assembly process with a conversion step at 2, 3, 4 or 5 monomers. (A, B): the dwell-time distributions for the first association step (A) and the first dissociation step (B) are shown. The dwell-time distribution for each individual step was fitted with a single-exponential distribution. (C, E): The extracted association rates for a nucleation (C) and a conversion mechanism (E) with different nucleation or conversion sizes. The dwell-time analysis is able to identify the association rates for the individual steps. The association rates during nucleation $k^+_{nuc}$ were set to 50% of $k^+_{poly}$ with no change in the off-rates. (D, F): The extracted dissociation rates for a nucleation (C) and a conversion mechanism (E) with different nucleation or conversion sizes. The dwell-time analysis could not identify the dissociation rate changes during the nucleation or conversion steps. The dissociation rates during nucleation $k^-_{nuc}$ were chosen to be 5x of $k^-_{poly}$ with no change in the on-rates. Error bars for the rates represent the 95% confidence intervals of the exponential fits.

During a nucleation process or around a conversion step, the self-assembling oligomers are likely to fluctuate around the rate-limiting steps until the nucleus size is reached or the necessary rearrangement has occurred. To quantify these dynamics, we calculated the number of upward and downward transitions that start from a certain monomer number (additional information provided by the visitation analysis). With this approach, a nucleation mechanism can be visualized (Figure S7G). However, the difference in the transitions around one single step is not sufficient for the detection of a conversion step with this approach (Figure S5E). Furthermore, the number of visits or the meantime per visit were calculated as an alternative to calculating the mean time the oligomers spent at a certain oligomer size. Both approaches could identify a change in the on- and off-rates of individual steps similar to the visitation analysis (Figure S5).

Since the visitation analysis could detect conversion steps with a high sensitivity, we wanted to further test its ability to identify single steps with different kinetic rates. Therefore, we simulated a growth process with individual slower or faster steps at defined oligomer sizes as depicted in Figure S5A. Here, the visitation analysis is sensitive to changes in the on and off-rates of individual steps with only slight differences in the rates (Figure S6). The dwell-time analysis could not identify even one of the faster or slower kinetic steps (Figure S5F). Thus, the visitation analysis could identify even multiple conversion steps during an assembly process.

A nucleation phase is characterized by a slower on-rate respective to the polymerization kinetics, or by a destabilization effect via high off-rates. These two mechanisms are visible in the average rates (Figure S4). A slower on-rate during nucleation results in a steady increase in the average rates with oligomer size (Figure S4A) whereas, with a faster off-rate, the average rates first decrease with oligomer size and then increase again (Figure S4B). Therefore, the average rates can be used to determine whether the association or the dissociation kinetics are changed when the nucleus size or conversion step is reached.

### Influence of SNR and data collection rate

Every analysis tool has its requirements and limitations depending on the experimental data. The requirements are coupled to the question that one wishes to address: do we want to determine the nucleation mechanism or is the aim to correctly determine every microscopic rate for each step? To estimate the required SNR and measurement rate for correctly answering these questions, we tested the presented analysis methods for their robustness as a function of SNR and sampling times. The average-rate analysis proved to be the most robust towards low SNR and slow data collection rates (Figure S2). The visitation analysis (Figure S6, Figure S7) and the dwell-time analysis (Figure S8) both use single-filament data and a step-finding algorithm. However, the visitation analysis is much more robust towards SNR and slow measurement times. In addition, it is able to detect multiple individual steps even at an SNR of 0.5 and a measurement rate of 5 times the fastest on-rate (Figure S6). The visitation analysis is also very robust towards SNR and the measurement rate when detecting a nucleation mechanism (Figure S7, table 1).

Using a multi-step correction, we could improve the performance of the dwell-time analysis also for measurements with a slow data collection rate (Figure S8F). The approach uses the intensity of the steps to identify single, double or triple steps (see materials and methods for details) and to correctly transform this information into the appropriate number of monomers, even when fast single steps cannot be resolved.

### Effect of labeling efficiency and photobleaching

Imaging of filament growth in ZMW or TIRFM relies on fluorescent labeling of the monomers. Labeling of the proteins for the nucleation studies brings additional complexities into the analysis. Dependent on whether a stochastic or a specific labeling strategy is used, the fraction of labeled monomers can either show a Poisson distribution of labels, where there may be more than one dye per monomer, or each monomer has either zero or one dye molecule attached as only one binding site is present when specific labeling approaches are used. Though high-labeling efficiencies are, in some cases, possible (*24*), sample preparation can be very time-intensive. At times, it is also necessary to use a mixture of labeled and unlabeled monomers to avoid influencing the assembly process. When the protein is not specifically labeled with 100% labeling efficiency, the labeling efficiency needs to be accounted for.

Proteins are typically labeled stochastically on naturally occurring lysine or cysteine residues, which results in a Poisson distribution of labels. In this case, one monomer may contain more than one dye molecule. When only one labeling cite is available, the monomers can be specifically labeling resulting in zero or one fluorophore per monomer. Hence, after simulating a monomer trace for the given conditions, we modified the trace assuming either a Poisson distribution of different labeling efficiencies from 0.3 to 3, or with a specific labeling approach with labeling efficiencies from 0.3 to 1 (Figure S9). For determining the nucleation mechanism, labeling efficiencies below 100% do not provide any difficulties for all three analysis methods (Figure S9A-F). For detecting single steps, however, the dwell-time analysis and visitation analysis require high labeling efficiencies (Figure S9G-M). Only the average-rate analysis can deal with labeling efficiencies down to 30% (Figure S9H, L). For the dwell-time and the average-rate analysis, the extracted rates for the polymerization regime correspond to $k^+_{poly}$ as LE* $k^+_{poly}$. Thus, if the process reaches polymerization kinetics and the labeling efficiency is known, the correct polymerization rate can be extracted quantitatively.

Another artifact that impacts the measurements is the photobleaching of fluorophores. Although photobleaching can be reduced using oxygen scavenging systems, it will always affect the intensity signal depending on the bleaching rate. In intensity traces, a photobleaching step is indistinguishable from a dissociation step. This hinders not only the direct measurement of the dissociation rate in single-filament traces but also leads to a mismatch between the intensity level and the monomer number, which influences the extracted on-rates as well. To test for the influence of photobleaching, we introduced down steps in the simulations based on different photobleaching rates. These rates are in the regime of experimentally determined values ((*6*), Figure S14). When a monomer dissociates after photobleaching, no down step is introduced. If not indicated otherwise, photobleaching was applied to 100% specifically labeled monomers.

With respect to distinguishing between different mechanisms of nucleation, the presented analysis methods are all affected by photobleaching, though some still correctly identify the underlying mechanism (table 1). For all analysis methods, photobleaching should not exceed 3% of the on-rate during polymerization or 30% of the off-rate (Figure S7, Figure S2, Figure S10). For the average rate analysis, the apparent average rates can decrease with oligomer size due to the effect of photobleaching. However, the signature of a nucleation or conversion mechanism is still visible at a photobleaching rate of 3% of the simulated on-rate (Figure S7 G, H). The visitation analysis of a growth process that assembles fast compared to photobleaching is largely unaffected (Figure S7, Figure S2).

Using the dwell-time analysis, the influence of photobleaching can also be used as a tool by comparing measurements with different photobleaching rates. In this way, it is possible to distinguish between a nucleation mechanism that affects the on-rate or the off-rate (Figure S9). A high off-rate leads to the exchange of photobleached monomers with unbleached monomers. By comparing the apparent off-rates of measurements with different photobleaching rates as determined by the dwell-time analysis, the nucleus size or a conversion step could be correctly determined even at higher photobleaching rates (Figure S10).

To visualize the combined effect of labeling efficiency and photobleaching on different oligomer growth scenarios, we simulated unhindered filament growth without nucleation or conversion and a filaments whose growth was restricted to a tetramer (Figure 3). A photobleaching rate at 10% of the polymerization kinetics still allows correct kinetic analysis via the visitation analysis, dwell-times analysis and average rates (Figure 3 red curves). In combination with a stochastic labeling efficiency of 30%, the visitation analysis shows a distribution shifted towards smaller oligomers sizes (Figure 3A, E). The dwell-time analysis resulted in reduced rates (Figure 3 B, F). For a restricted growth mechanism where the growth stops upon obtaining a certain oligomer size (exemplified for a tetramer in Figure 3E-H), the average rates can only be extracted for the very first monomers because the average filament size does not reach beyond 0.5 monomers (Figure 3 G). For that reason, fractional average rates in 0.1 monomer steps have been extracted (Figure 3 H).

### Experimental Results

We used the three analysis methods (the dwell-time analysis, visitation analysis, and fractional average-rate analysis) to investigate the nucleation mechanism of the actin nucleators Cappuccino and the ABCD fragment of Spire. The formin Cappuccino stabilizes actin monomers via its FH2 domains, thereby promoting nucleation (*25-27*). Spire-ABCD contains 4 Wiskott–Aldrich syndrome protein (WASP) homology 2 (WH2) domains, labeled as ABCD, which bind actin monomers (*28, 29*). For the experiments, 30% stochastically labeled G-actin-Cy5 was used. The biotinylated nucleator proteins were attached to the bottom of zero-mode waveguides via a biotin-streptavidin interaction, and the actin monomers were added directly after the start of the measurement (see materials and methods and ref (*6*) for details). Actin-Cy5, as well as biotinylated Cappuccino, were fully functional (Figure S 11, Figure S12).

For actin growth on the strong nucleator Cappuccino (Figure 4A-D), we observed an increase in the average filament size over time (Figure 4B) reaching oligomer sizes of up to 10 monomers already in the first 400 s (Figure 4A). The dwell-time analysis showed increasing rates with each step (Figure 4D). The fractional average rate analysis, however, shows a decrease in the association rates with oligomer size. This observation can be explained by the influence of a labeling efficiency of 30% and photobleaching. Thus, even a simulation of unhindered growth, i.e. no nucleation or conversion mechanism, under these conditions showed decreasing average rates with oligomer size (Figure 3C).

Measurements of actin nucleation in the presence of Spire-ABCD (Figure 4E-H) did not show prominent growth in the average filament trace (Figure 4F), even though the dwell-time analysis showed increasing on-rates with oligomer size (Figure 4H). In contrast, the visitation analysis, showed that only very few monomers bind to Spire-ABCD (Figure 4E). The discrepancy between the dwell-time analysis and the other two approaches, i.e., visitation analysis and average-rate analysis, can be explained by a bias in the on-rates. When filament growth is very unlikely, only fast on-rates can lead to the assembly of higher oligomers, thus filtering the distribution of individual on-rates for the fast rates. This affects the dwell-time analysis, but not the visitation and average-rate analysis. Comparing the resulting actin growth on Spire-ABCD with simulated data affected by photobleaching and labeling efficiency, a restricted growth model up to a few monomers could reasonable describe the data (Figure 3E-H).

## Discussion

Single filament data contain information regarding the nucleation mechanism that is not immediately visible from inspection of the recorded time traces. The most common approach for dealing with data showing the individual association and dissociation steps is the kinetic analysis of the dwell-times. The obtained microscopic rates should directly reflect the underlying self-assembly mechanism. However, the correct interpretation of dwell-time distributions depends, to a high extend, on the quality of the data (i.e. signal-to-noise ratio, SNR) and also on the measurement rate or the influence of dye photobleaching in fluorescence microscopy measurements (Table 1). Therefore, we developed more robust analysis methods that give insights into the microscopic mechanism of the self-assembly process and are less prone to misinterpretation. An overview of the minimal requirements for each of the tested analysis methods can be found in Table 1 - with respect to the measurement rate, SNR, labeling efficiency and photobleaching rate.

The dwell-time analysis provides the microscopic rates for the addition of each monomer and should, therefore, reveal the underlying assembly mechanism. However, it is very sensitive to the SNR and acquisition speed, which needs to be more than an order of magnitude faster than the expected kinetics (Table 1). Also, photobleaching leads very quickly to a loss of synchronization between the measured and actual filament size. However, one can also apply photobleaching as a tool by measuring filament formation with different photobleaching rates. A comparison of the extracted rates can help to identify the nucleus size or conversion site (Figure S10).

The visitation analysis is able to detect a slow nucleation phase despite non-ideal measurement conditions like low SNR (Figure S7). Moreover, it is very suitable to identify multiple events with slower or faster kinetics during an assembly process (Figure S6). For the analysis, the sampling rate needs to be about a factor of 2 higher than the expected kinetics, which is an order of magnitude slower than the required speed for the dwell-time analysis (Table 1). The visitation analysis on single-filament data is, therefore, a suitable method for detecting nucleation or individual conversion steps (Table 1).

When the quality of single-filament data is not sufficient for the use of a step-finding algorithm to extract the individual monomer binding and dissociation events, an average trace can be obtained from the growth information of multiple filaments. When the number of filaments and the intensity information of a single monomer binding event is known, average rates can be used to visualize and detect a slow nucleation phase or a pronounced slow conversion step. The average-rate analysis can deal with a sampling time in the same range as the expected kinetics and a SNR as low as 0.1 (Figure S2, Table 1). Therefore, it is the most robust analysis method present here.

For the simulations of a nucleation or conversion process, changes in the dissociation and the association rates have been separated to study the effects independently. When only the association rate or the dissociation rate is affected during the growth process, the average rate analysis can be used to determine which rate changes when the nucleus size is reached (Figure S4).

Photobleaching, in general, affects all analysis methods (Table 1). The dwell-time analysis is most affected by photobleaching as the kinetics of single steps has to be determined (Figure S8, Table 1). The visitation analysis and the average-rate analysis require a photobleaching rate that should not exceed 2 to 3% of the association rate (Table 1). The labeling efficiency should be at least 30 to 50%, depending on the analysis method used and whether stochastic or specific labeling is used (Figure S9). The visitation analysis allowed for labeling efficiencies down to 30% for the detection of a nucleation mechanism (Figure S7, Figure S9). In contrast, the dwell-time analysis reproduced the correct on-rate only with very high labeling efficiencies (Figure S8 and Figure S9). In the case of specific labeling of the monomers, a nucleation process could still be identified even at 30% labeling efficiency with all three presented methods. A specific conversion-step could only be detected with the correct oligomer size for a specific labeling efficiency of 100%. However, the existence of a conversion step without a clear indication of the oligomer size could be detected at lower labeling efficiencies (Figure S9).

With analyzing experimental data, such as actin growth on Spire-ABCD and Cappuccino shown here, the experimental details like sampling rate, labeling efficiency and photobleaching have to be taken into account. To check the influence of the sampling rate on the dwell-time analysis experimentally - as carried out in the simulations (Figure S8), we increased the sampling rate for the experiments with Cappuccino from 5 Hz to 16.6 Hz. This led to an increase of the estimated association rate (0.24 $s^{-1}$ or 0.72 $s^{-1}$ when accounting for the labeling efficiency, Figure S13C). However, the extracted rates are still slower than the expected 8 $s^{-1}$ for 800 nM actin (*23, 30, 31*). For association rates on the order of 10 $\mu M^{-1}s^{-1}$ and a labeling efficiency of 30%, the necessary sampling rate for the dwell-time analysis is at 60 $s^{-1}$ (20*0.3*10 $s^{-1}$ = 60 $s^{-1}$), as the dwell-time analysis yields trustworthy rates at >20 times the expected association rate (Figure S8). These results suggest that the current time resolution used in these experiments is insufficient for an accurate estimation of the association rates via a dwell-time analysis. The visitation analysis and the average rates, however, clearly identify fast actin assembly as expected (Figure 4).

The photobleaching rate of actin-Cy5 showed a double exponential decay as expected for Cy5 (*32*). The slower rate was at 0.025 $s^{-1}$, which corresponds to 3.5% of the extracted on-rate corrected for the labeling efficiency (0.72 $s^{-1}$). That corresponds to the highest photobleaching rate used here in simulations (3% of the on rate). For the average-rate analysis, a photobleaching rate of 3% results in a decrease and even a cut-off of the extracted rates (Figure 3). This explains the decrease of the average rates for actin-Cy5 growing on Cappuccino (Figure 4).

In contrast to Cappuccino, where dwell-time analysis, visitation analysis, and average-rate analysis indicated continuous growth, actin on Spire-ABCD only formed smaller oligomers (Figure 4). This cannot be explained by photobleaching or other experimental settings. Therefore, Spire-ABCD can bind actin monomers but is insufficient to promote nucleation from monomers alone, as has been suggested before (*33*).

In summary, we developed new analysis methods for single and averaged filament data that can be obtained by fluorescence microscopy or other methods like iScat (*12*). These analysis methods are highly robust towards low SNR and slow measurement rates, in contrast to the widely used dwell-time analysis. These methods will help to elucidate the very first stages in filament formation.

# Materials and Methods

## Analysis

The noise-overlaid results of the simulations and the intensity measurements of the experimental data have been fed into a step-finding algorithm (*34*). The position in time and the change in intensity associated with monomer addition or dissociation (steps) were identified by the Salapaka step-finding algorithm (*34*) To prevent overfitting, the algorithm uses a penalty factor for the introduction of new steps (*34*). The results of the step finding algorithm were converted into a monomer number versus time trace. Thereby, an up step was assumed to correspond to one binding monomer, a down step to one dissociating monomer.

### *Dwell-time analysis*

For comparison to our newly developed analysis tools, we investigated the distribution of dwell-times to calculate the stochastic rate constant from the fluorescence trajectory. For the dwell-time analysis at slow sampling rates, a double step correction was used. With slow sampling, fast individual steps cannot be resolved and appear as a single step with higher step sizes. For the double step correction, the mean step size per trace was calculated. An up or down step with a step size twice or three times the mean step size was treated as two or three monomers, respectively. The additional steps were assigned a dwell-time that corresponded to the interframe time of the measurement. This caused higher numbers in the first bin of the dwell-time histogram (Figure S13). Therefore, the first bin was not included in the exponential fit.

## Experimental data

### Labeled Actin

Cy5-labeled actin (rabbit, skeletal muscle) was purchased from Hypermol (Bielefeld, Germany). We showed that fluorescently labeled actin is fully active as assayed by bulk techniques, TIRFM and single molecule methods (*35*).

### Expression and purification of Cappuccino and Spire constructs

Constructs encoding Spire-ABCD and Cappuccino, GST-ABCD1 and GST-CapuFH2, were transformed in E. coli BL21-CodonPlus-RIL competent cells and cultured in LB medium with ampicillin (100 µg/mL) and chloramphenicol (34 µg/mL) at 37°C. Protein expression was induced with 0.5 mM isopropyl-β-D-thiogalactopyranoside (IPTG) when OD600 reached 0.6-0.8 value.

Proteins were expressed for 16-20 h at 16°C. Cells were harvested by centrifugation (30 min, 2000 rpm, 4°C) and bacterial pellets were frozen at -20°C.

Subsequently, pellets were handled on ice using ice-cold buffers. Pellets were thawed on ice and resuspended in ca. 50 mL of lysis buffer A (1xPBS, pH 7.4, 1 mM EDTA, 1 mM DTT, 0.5 mM PMSF) in the case of GST-ABCD1 or lysis buffer B (50 mM Tris-HCl pH 7.0, 150 mM NaCl, 0.2% Triton-X100, 1 mM DTT, 1 mM PMSF and 1 µg/mL DNAseI) in the case of GST-CapuFH2. After cell disruption by ultrasonification, the whole extract was clarified by centrifugation (30 min, 15 000 rpm, 4°C). The soluble fraction containing recombinant proteins was then loaded onto a GSTPrep FF 16/10 column (GE Healthcare) connected to an ÄKTA FPLC System (GE Healthcare). After removal of bacterial host proteins, the column was washed with 10 column volumes of PBS and then equilibrated with PreScission Protease buffer (50 mM Tris-HCl, pH 7.0, 150 mM NaCl, 1 mM EDTA, 1 mM DTT). Next, a solution of PreScission Protease (4 mg/mL) was loaded onto the column and incubated overnight at 4°C. Subsequently, cleaved proteins were eluted from the column with PreScission Protease Buffer. The GST tag that remained on the column and PreScission Protease were removed by washing the column with regeneration buffer (50 mM Tris-HCl pH 8.0, 150 mM NaCl, 10 mM GSH). Fractions containing ABCD1 or CapuFH2 were concentrated and subsequently purified by a gel filtration method on S75 Superdex column (ABCD1) or on a S200 Superdex column (CapuFH2) that were equilibrated with storage buffer (50 mM Tris-HCl, pH 8.0, 300 mM NaCl, 1 mM DTT).

Site-specific biotinylation via Sortase A-mediated ligation (SML)

Spire-ABCD and Cappuccino constructs were labelled at their N-termini using Sortase A-mediated ligation. In a typical reaction, CapuFH2 (45 µM) or ABCD1 (50 µM) were mixed with an excess of desthiobiotin-peptide (300 µM) bearing a sequence recognized by Sortase A (desthiobiotin-GCGLPETGG, Smart Bioscience) and Sortase A (2 µM, Cat. #E4400-01; Eurx). The reaction mixture was supplemented with 10 mM $CaCl_2$ and incubated for 6 or 24 h at 4°C (in the case of ABCD1) or at 4°C and 32°C (for CapuFH2). The progress of the reaction was monitored with immunoblotting using HRP-conjugated streptavidin (Cat. #405210; BioLegend). After completion of the desthiobiotynilation reaction, both proteins were purified by size-exclusion chromatography. Desthiobiotin-CapuFH2 (DB-CapuFH2) was purified on a Superdex 200 Increase 10/300 column (GE Healthcare). The Desthiobiotin-ABCD1 construct was purified on a Superdex S75 column. The degree of desthiobiotinylation was calculated to be 53% and 60% for CapuFH2 and ABCD1, respectively as judged by an HABA assay (Pierce Biotin Quantification Kit, ThermoFisher Scientific).

### Bulk assays

The biological activity of (DB-)CapuFH2 and (DB-)ABCD1, as well as functional actin growth were verified in bulk. Pyrene actin polymerization assays were done using 10% pyrene-labelled actin (C374) (PA, Hypermol) in black 96-well plates (Corning). Before the measurements, actin dissolved in G-buffer (Cytoskeleton) was incubated in ME-buffer (50 µM $MgCl_2$, 0.2 mM EGTA) for 5 min on ice. 80 µL of a 5.0 µM actin solution were placed in selected wells of the plate. Actin polymerization was induced by adding 20 µL of CapuFH2 or DB-CapuFH2 in KMEI buffer (final concentrations: 5 mM KCl, 1 mM $MgCl_2$, 0.5 mM EGTA and 0.2 mM imidazole). The final concentration of actin was 4 µM. The pyrene-fluorescence was excited at 350 nm and emission was measured at 520 nm on a TECAN microplate reader.

### ZMW fabrication and functionalization

ZMWs were fabricated at the Imec in Leuven, Belgium. In short, ZMWs were fabricated on glass coverslips (22mmx22mm, type #1, Menzel Glazer) by means of physical vapor deposition, e-beam lithography and low-pressure dry etching. ZMWs were passivated using 0.2% polyvinylphosphonic acid (Polysciences, USA) (10 min, 90°C), followed by incubation in 3-[Methoxy(polyethyleneoxy)propyl]trimethoxysilane (6-9 PE-units, abcr, Karlsruhe, Germany) and PEG-biotin-silane (Nanocs Inc, New York, USA) in toluene (4h, 55°C).

### Fluorescence-based microscopy and data extraction.

Experiments were carried out on a home-built wide-field microscope system equipped with a 60x 1.45 NA oil immersion objective (Plan Apo TIRF 60x; Nikon). The laser power of the 642 nm laser (Cobolt MLD; Cobolt AB) was set to 2.8 mW before entering the objective via a dichroic mirror (zt405/488/561/640rpc; AHF Analysentechnik, Tübingen, Germany). The fluorescence signal was collected using the same objective, passing through an emission filter (680/42 BrightLineHC; AHF Analysentechnik, Tübingen, Germany) and recorded on an EMCCD camera (Andor iXon Ultra 888; Andor Technology) using an integration time of 50 ms.

The intensity as a function of time per aperture was extracted using a custom written software in MATLAB (The MathWorks) as described previously (*6*). In short, the fluorescence signal of a signal mask matching the apertures was converted into an intensity versus time trace. After finding the up and down steps via a step-finding algorithm (*34*), the intensity trace was converted into a monomer number trace (see (*6*) for details), which was then used for further analysis.

### Single-molecule Imaging of Actin Polymerization in Presence of Nucleators

Actin growth on Cappuccino and Spire-ABCD was measured in ZMWs. ZMWs were incubated with 0.15 mg/ml Streptavidin and 5 mg/ml BSA in PBS for 5 min. After washing with PBS, 30 nM biotinylated Cappuccino or Spire-ABCD was incubated for 5 min, followed by 5 min incubation with 1 mg/ml BSA and 1 mg/ml biotinylated BSA to block free streptavidin molecules on the surface. Actin was prepared as described in the following paragraph and added after washing with PBS. Actin-Cy5 in G-buffer was incubated in magnesium exchange (ME) buffer for 5 minutes on ice (50 µM $MgCl_2$, 0.2 mM EGTA, pH 7.0). Actin polymerization was induced by adding 1/10 volume of a 10x concentrated KMEI buffer (final concentrations: 50 mM KCl, 1 mM $MgCl_2$, 0.5 mM EGTA, and 0.2 mM imidazole at pH 7.0). The final buffer contained a PCA/PCD oxygen scavenging system with 250 nM protocatechuate dioxygenase 'PCD', 2.5 mM 3,4-dihydroxybenzoic acid 'PCA' and 1 mM Trolox (*36*). After starting polymerization, 50 µl of the reaction mixture was added to the waveguides and data acquisition was started immediately with a measurement rate of 5 Hz, if not indicated otherwise. To measure the photobleaching rate, 1 µl of phalloidin was added to the waveguides after 1.5 h to stabilize the formed filaments and prevent dissociation.

### TIRF imaging of actin filaments

TIRF microscopy of actin filaments was performed on the same setup as described above. The actin was treated in the same way as for the ZMW experiments, but instead of adding the solution onto the waveguides, a flow cell system was used as described in (*31*). The length of the filaments was analyzed 250 s after induction of polymerization to verify the full functionality of actin-Cy5 Figure S1Figure S 11).

## Acknowledgements

This work was supported by a grant from the Deutsche Forschungsgemeinschaft through the SPP 1464 (to AHC and DCL), the Excellent Clusters Nanosystems Initative Munich (NIM) and Center for Integrated Protein Science Munich (CIPSM), and the Ludwig-Maximilians-Universität München (via the LMUInnovativ BioImaging Network (BIN) and the Center for NanoScience (CeNS)). Work in the Crevenna laboratory was financially supported by: Project LISBOA-01-0145-FEDER-007660 (Microbiologia Molecular, Estrutural e Celular) funded by FEDER funds through COMPETE2020 - Programa Operacional Competitividade e Internacionalização (POCI) and by national funds through FCT - Fundação para a Ciência e a Tecnologia. This research has also been supported by the 2011/01/B/NZ1/00031 grant (to TAH) from the National Science Centre, Poland. RK received financial support (doctoral scholarship) within the ETIUDA Programme from the National Science Centre, Poland (Project No: UMO-2017/24/T/NZ1/00272 and UMO-2017/27/N/ST5/01405).

Table 1: Robustness of the analysis methods

|  | Dwell-time analysis | | Visitation analysis | | Average-rate analysis | |
| --- | --- | --- | --- | --- | --- | --- |
|  | Mechanism | Single steps | Mechanism | Single steps | Mechanism | Single steps |
| *SNR* | >1 | >1 | >0.1 | >1 | >0.1 | >0.1 |
| *Sampling rate* | >10x* | >20x | >1x | >2x | >1x | >5x |
| *LE (stochastic)* | >1 | >2 | <1 | 1 | >0.3 | >0.7(1**) |
| *LE (specific)* | 1 | >0.3 | 1 | >0.3 | 1 | >0.3 |
| *Photobleaching* | <3%* | <0.1% | 3% | <2% | <3% | <2% |

\* compared to the fastest rate
\*\* 1 to obtain the correct conversion site

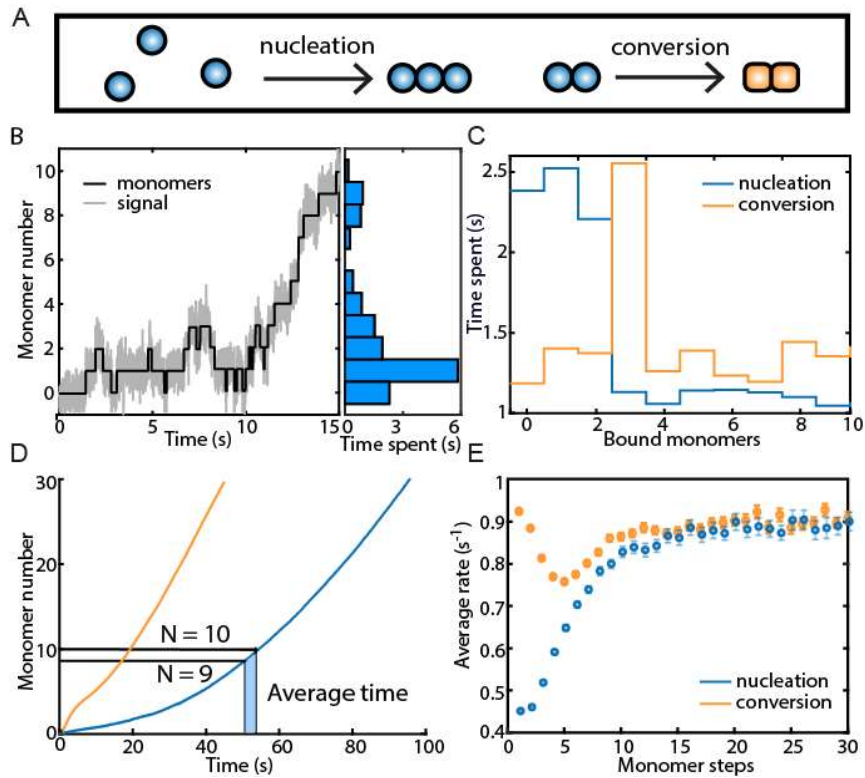

Figure 1: Analysis methods for single-filament data of protein self-assembly. A: B: Example trace of a simulated growth process undergoing a nucleation mechanism with a nucleus size of four monomers (black) with Gaussian noise (grey). The distribution of time spent at a certain oligomer size (blue histogram) contains information about the nucleation mechanism when it is applied to many single oligomer traces (C). C: Visitation analysis on a nucleation (blue) or a conversion mechanism (yellow). The visitation analysis identifies the nucleation mechanism as well as the nucleus size for nucleation or conversion. D, E: Average rates analysis. D: average traces of a nucleation (blue) or conversion (yellow) mechanism. From the average trace of many individual filaments growing from the same starting point, average rates can be calculated from the average time it takes to reach always the next oligomer size. E: average rates plotted versus the monomer number (oligomer size). Different nucleation mechanisms show individual signatures (E). The nucleation and conversion kinetics were simulated to show slower association rates than the polymerization kinetics.

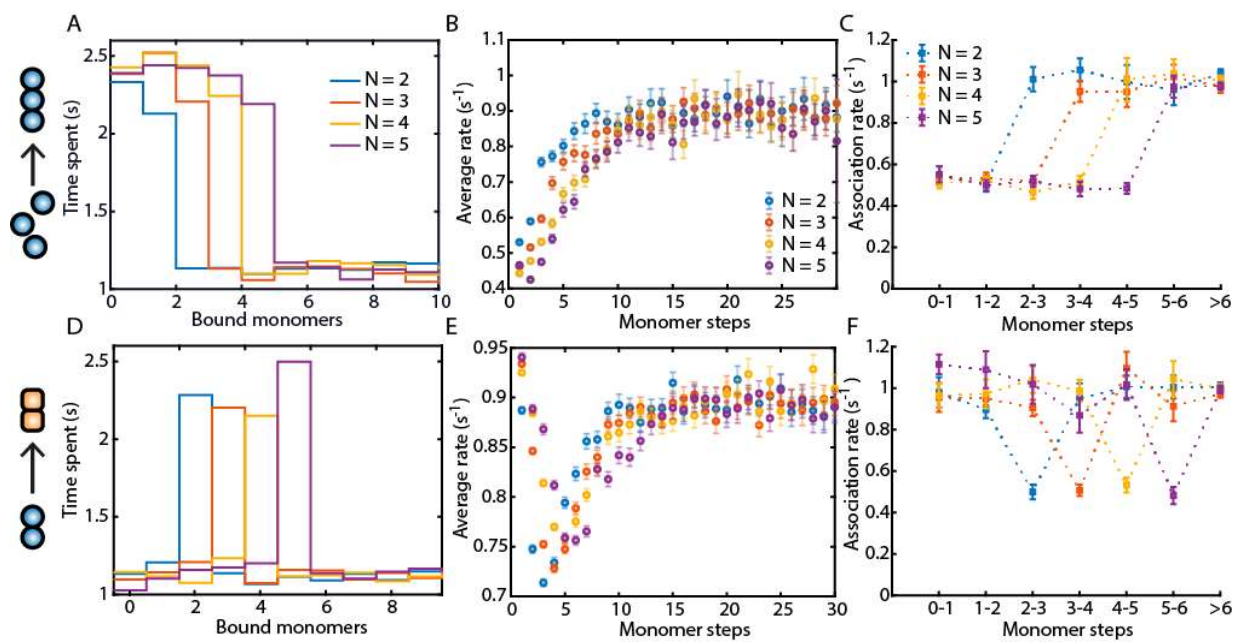

Figure 2: A comparison of the visitation analysis, average rates and dwell-time analysis on an assembly process with nucleus or conversion size of 2, 3, 4 or 5 monomers. (A, D): A visitation analysis on a nucleation (A) and on a conversion mechanism (D). (B, E): Average rate analysis on the same nucleation and conversion mechanism as in A and D. (C, F): Dwell-time analysis on the same nucleation and conversion mechanism as in A and D. The dwell-time analysis can identify a mechanism with a change in the on-rate by extracting the association rates (F), but not a mechanism with a change in the off-rate, since the extracted dissociation rates do not indicate any changes in the simulated rates (see Figure S3). Error bars for the rates represent the 95% confidence intervals of the exponential fits. The association rates during nucleation $k^+_{nuc}$ or the conversion step $k^+_{conv}$ were chosen to be 50% of $k^+_{poly}$ with no change in the off-rates.

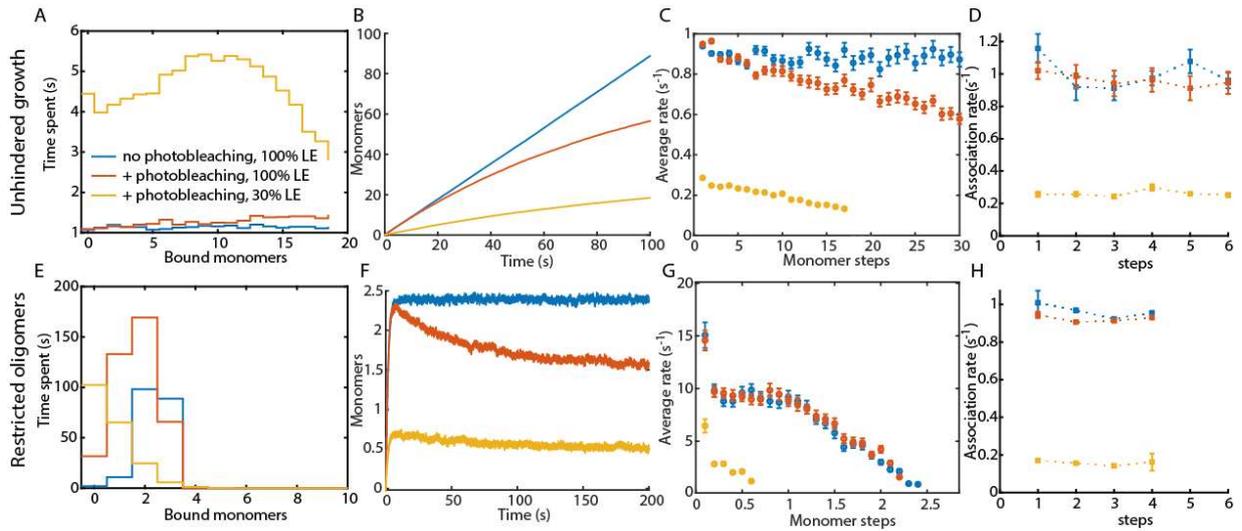

Figure 3: Effect of photobleaching and labeling efficiency on unhindered growth and restricted growth that stops after the assembly of four monomers. Both unhindered (A-D) and restricted growth (E-H) had $k^+_{poly}$ = 1s$^{-1}$ and $k^-_{poly}$ = 0.1s$^{-1}$. Visitation analysis (A, E), average rates (C, G) extracted from the average traces (B, F) and dwell-time analysis (D, H) were applied to data with 100% labeling efficiency and no photobleaching (blue data). The average trace of restricted growth reaches only 2.5 monomers, despite possible growth until 4 monomers, because of the equilibrium between the on- and the off-rate. The effect of a photobleaching rate of 0.01 s$^{-1}$ does not significantly influence the extracted rates and the visitation analysis (red data). The combination of stochastic labeling of 30% and photobleaching with a rate of 0.01 s$^{-1}$ affects all analysis methods (yellow). Error bars for the dwell-time analysis represent 95% confidence intervals of the exponential fit.

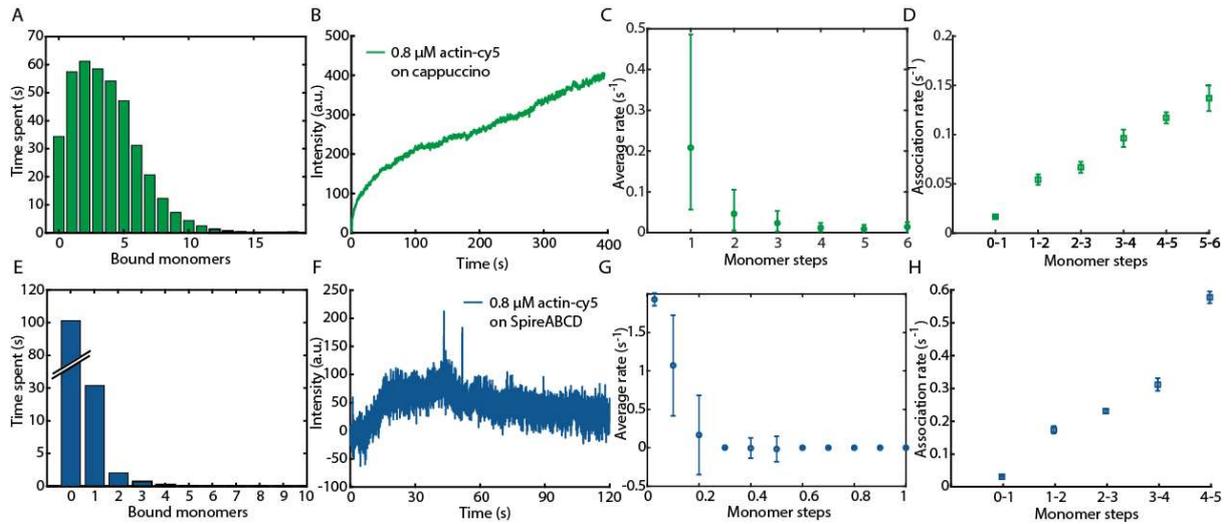

Figure 4: Experimental data of 30% stochastically labeled actin-Cy5 growing on Cappuccino and Spire-ABCD. A comparison between a visitation analysis, dwell-time analysis and average rate analysis is performed for the measurements with Cappuccino (A-D, green) and Spire-ABCD (E-G, blue) . (A, E): The visitation analysis, (B, F) average intensity, (C, G) fractional average-rate analysis and (D, H) dwell-time analysis showing the first 5 monomer association steps from 700 individual traces. The fractional average rates have been calculated using the time until the average trace reached the intensity change corresponding to 0.1 monomers (i.e. between 0 and 0.1, between 0.1 and 0.2, between 0.2 and 0.3, etc). Fractional average rates are 10 times faster than average rates because the average intensity of only 1/10 of a monomer has to be reached. The experimental photobleaching rate was 0.025 $s^{-1}$.

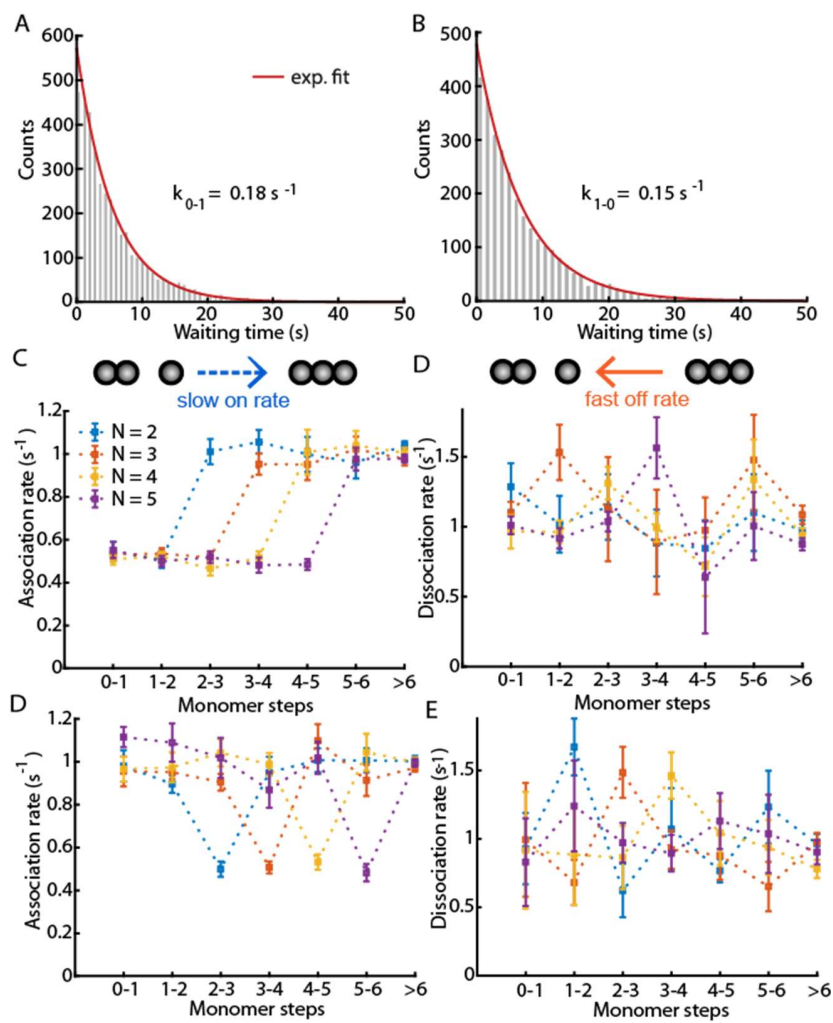

Figure S1: Dwell-time analysis on an assembly process with a conversion step at 2, 3, 4 or 5 monomers. (A, B): the dwell-time distributions for the first association step (A) and the first dissociation step (B) are shown. The dwell-time distribution for each individual step was fitted with a single-exponential distribution. (C, E): The extracted association rates for a nucleation (C) and a conversion mechanism (E) with different nucleation or conversion sizes. The dwell-time analysis is able to identify the association rates for the individual steps. The association rates during nucleation $k^+_{nuc}$ were set to 50% of $k^+_{poly}$ with no change in the off-rates. (D, F): The extracted dissociation rates for a nucleation (C) and a conversion mechanism (E) with different nucleation or conversion sizes. The dwell-time analysis could not identify the dissociation rate changes during the nucleation or conversion steps. The dissociation rates during nucleation $k^-_{nuc}$ were chosen to be 5x of $k^-_{poly}$ with no change in the on-rates. Error bars for the rates represent the 95% confidence intervals of the exponential fits.

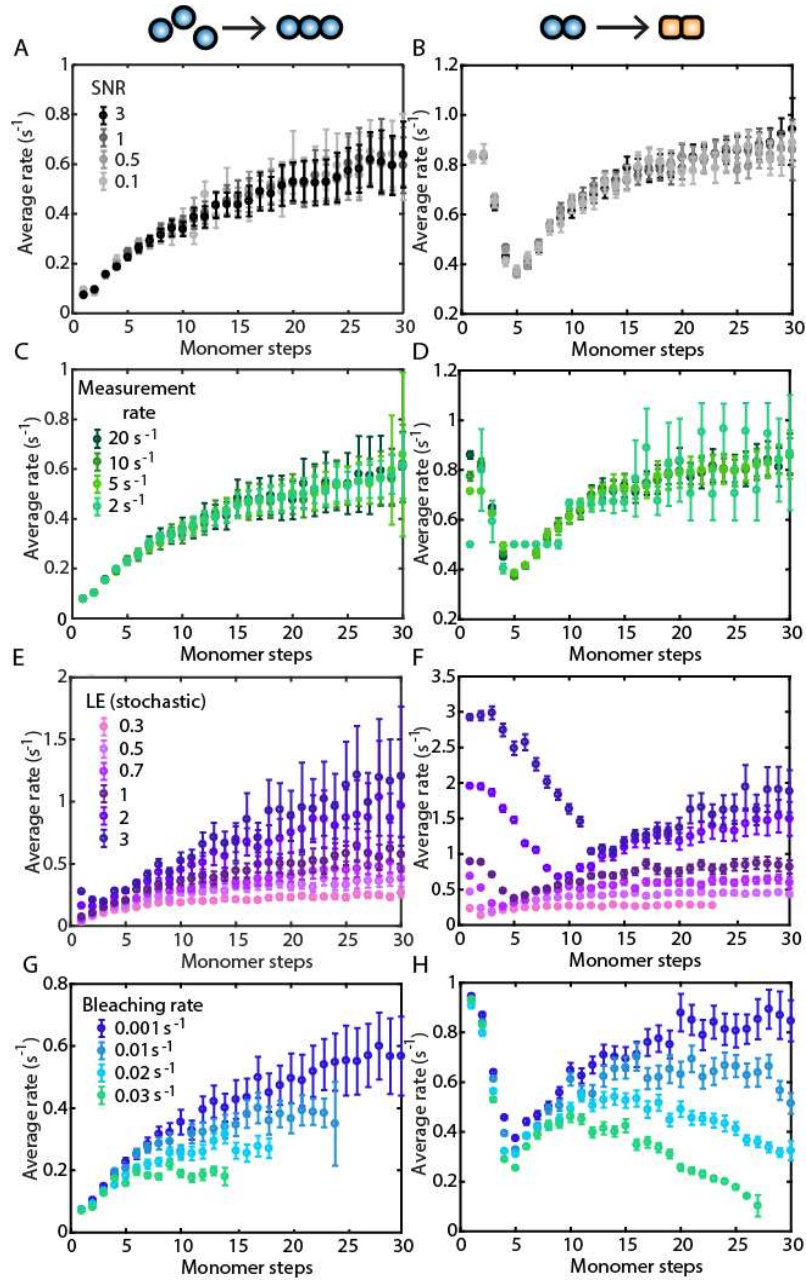

Figure S2: Robustness of the average rate analysis. (A, B): An average rate analysis on simulations applying different SNRs. A nucleation mechanism (A) or a slow conversion step (B) can be detected even at a SNR of 0.1. (C, D): An average rate analysis on simulations varying the sampling time. To detect a nucleation mechanism or conversion step, the measurement rates should be at least 2x faster than the fastest rate. (E, F): An average rate analysis for simulations using monomers with different labeling efficiencies. A nucleation mechanism (E) or a conversion step (F) can be detected at all tested labeling efficiencies down to 30%. For the correct conversion step (F), the labeling efficiency should be at 100%. (G, H): An average rate analysis for simulations considering the influence of photobleaching. A high photobleaching rate can shorten the average trace or lead to apparently slower kinetics.

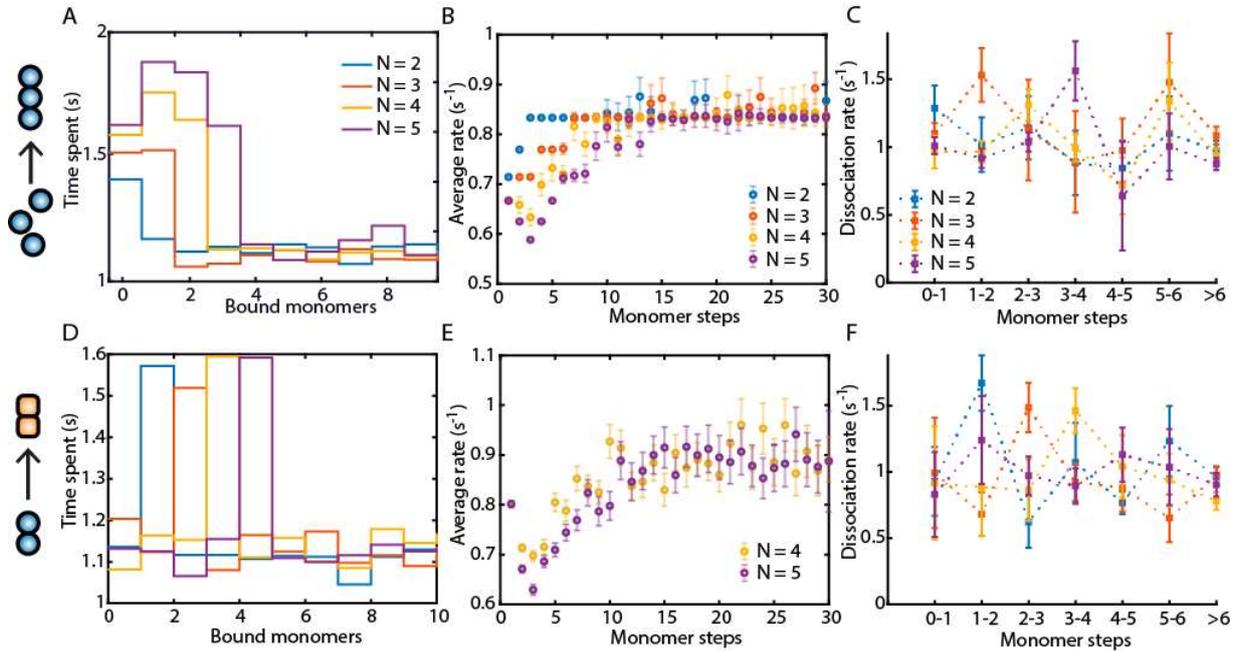

Figure S3: A comparison of the visitation analysis, average rates and dwell-time analysis on an assembly process with nucleus or conversion size of 2, 3, 4 or 5 monomers. The dissociation rates during nucleation $k^-_{nuc}$ were chosen to be 5x faster than $k^-_{poly}$ with no change in the on-rates. (A, D): A visitation analysis on a nucleation (A) and on a conversion mechanism (D). (B, E): Average rate analysis on the same nucleation and conversion mechanism as in (A, B). (C, F): A dwell-time analysis on the same nucleation and conversion mechanism as in (A, B). The dwell-time analysis cannot identify a mechanism with a change in the off-rate, since the extracted dissociation rates do not indicate any changes in the simulated rates. Error bars for the rates represent the 95% confidence intervals of the exponential fits.

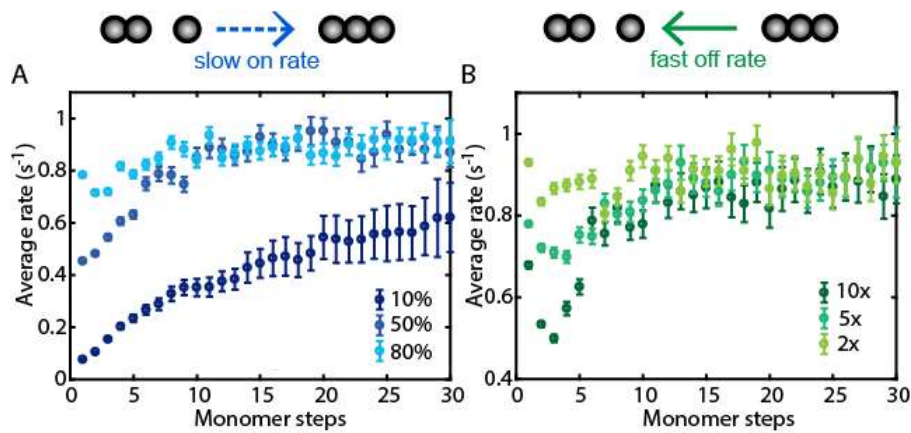

Figure S4: The influence of a change in the on-rate versus a change in the off-rate in the average rate analysis for a nucleation mechanism with a nucleus size of 4 monomers. (A): Average rate analysis on simulations varying the association rate during nucleation. The association rates during nucleation $k^+_{nuc}$ were set to 10, 50 or 80% of $k^+_{poly}$. (B): Average rate analysis on simulations varying the dissociation rate during nucleation. The dissociation rates during nucleation $k^-_{nuc}$ were a factor of 10, 5 or 2 of $k^-_{poly}$. The average rates show a different behavior dependent on whether the association rate or the dissociation rate was changed with respect to the polymerization rates.

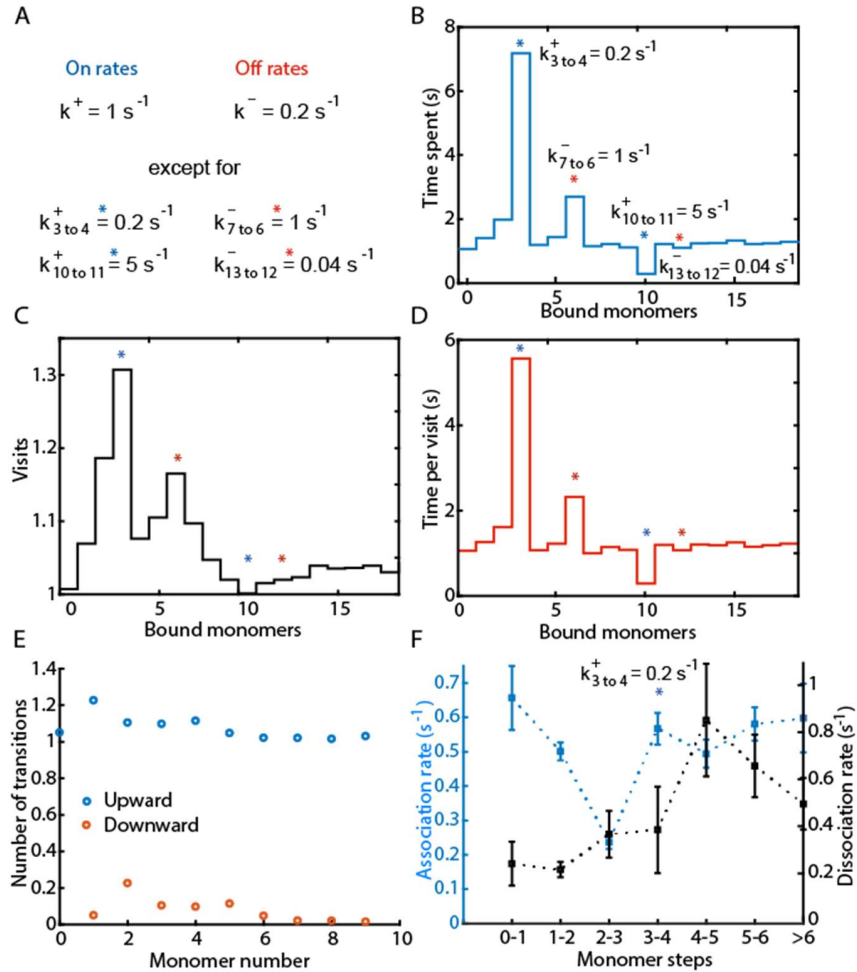

Figure S5: A visitation analysis on a growth process where single steps have different kinetics. (A): Model used for the simulation. All monomer binding events have an on-rate $k^+ = 1\ s^{-1}$, all dissociation events an off-rate $k^- = 0.2\ s^{-1}$ with the following exceptions: The third and 10th monomer binding event show a slower or a faster on-rate, respectively (by a factor of five, if not otherwise indicated) and are marked by blue stars; the 7th and 13th monomer binding event show a faster or slower off-rate, respectively, and are marked by red stars. (B-D): The visitation analysis can resolve multiple single steps with different kinetics. (B): The mean time the traces spent at a certain oligomer size is plotted. (C): The mean number of occurrences of a certain oligomer size (i.e., number of visits). (D): The mean time per visit at a certain oligomer size. (E-F): Multiple individual rates cannot be resolved by dwell-time analysis and number of transitions. (E): Mean number of upward (blue) and downward (red) transitions from a certain oligomer size (monomer number). (F): Dwell-time analysis on the same simulated growth process. The dwell-time analysis is not able to resolve multiple individual rates even with a SNR of 2, measurement rate of 20 $s^{-1}$ and no photobleaching.

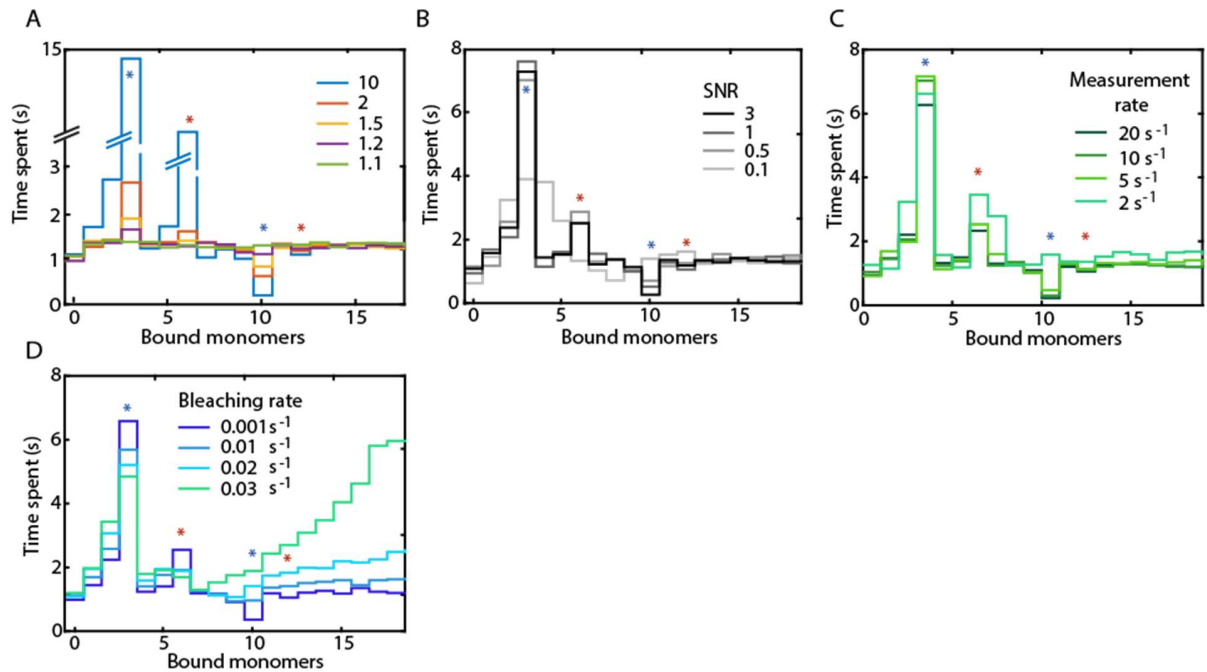

Figure S6: Reliability of the visitation analysis for a hypothetic growth process. (A): A visitation analysis of simulations using the model described in Figure S5A, where the varying the default rates by different factors. The difference between the single steps with slower or faster kinetics can still be detected for differences of a factor of 1.5. (B): Impact of the SNR. Using a difference in kinetic rates of a factor of five, we simulated SNR ratios from 3 to 0.1. Individual steps are visible down to a SNR of 0.5. (C): Varying the sampling rate. Using the model described in Figure S5A, we performed simulations with different sampling times. The visitation analysis is able to detect individual steps at measurement rates down to 5 times the default on-rate. (D): Visitation analysis on simulations considering the influence of photobleaching. The association rates during nucleation need to be fast compared to the photobleaching rate in order to resolve the first steps in an assembly process.

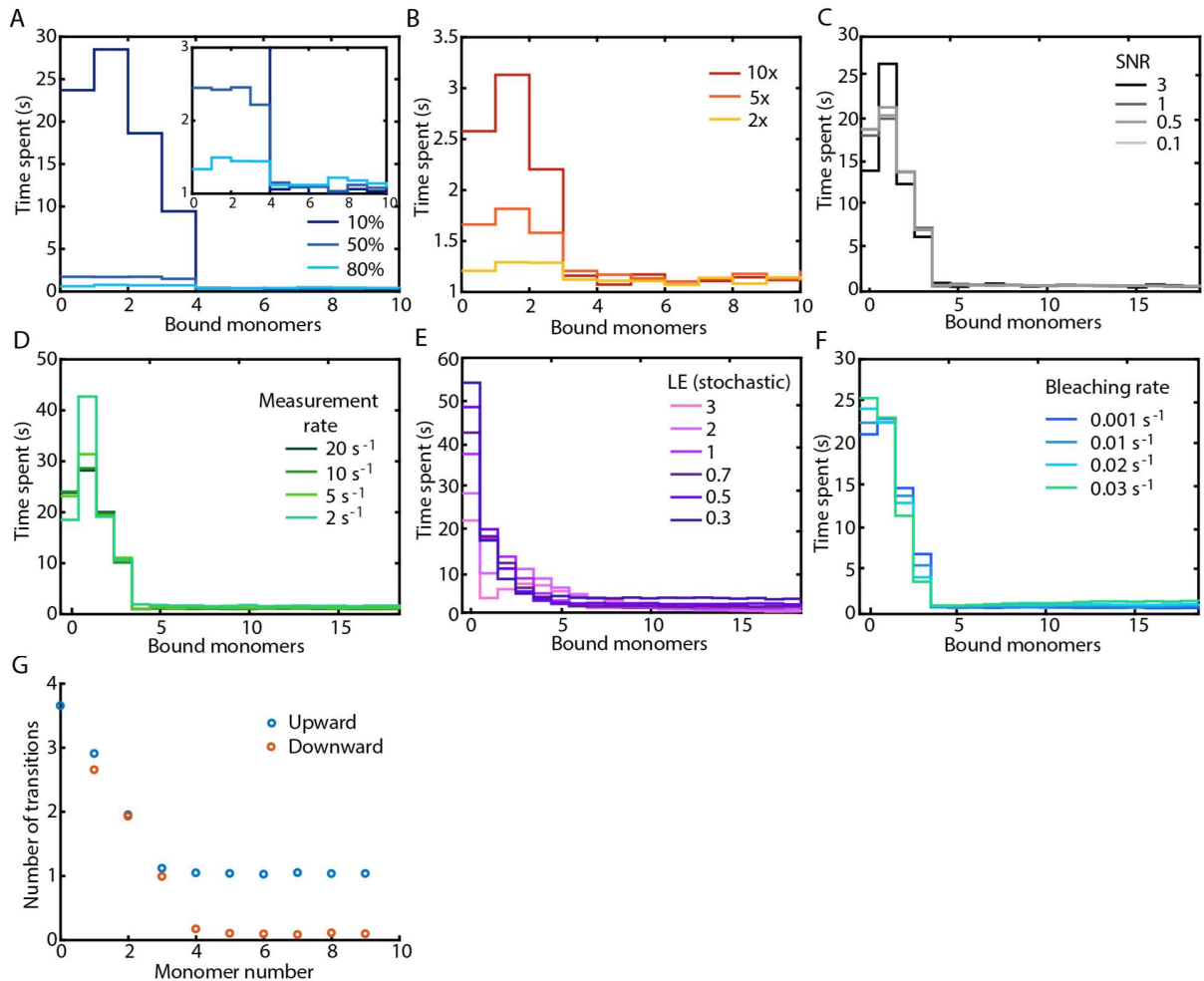

Figure S7: Robustness of the visitation analysis applied to a nucleation mechanism with a nucleus size of four monomers. (A, B): Difference between nucleation and polymerization kinetics. The association rates during nucleation $k^+_{nuc}$ were chosen to be 10, 50 or 80% of $k^+_{poly}$ (A). The dissociation rates during nucleation $k^-_{nuc}$ were a factor of 10, 5 or 2 of $k^-_{poly}$ (B). (C): Visitation analysis on simulations varying the SNR. A nucleation mechanism can be detected until a SNR of 0.1. (D): Visitation analysis on simulations varying the sampling time. The visitation analysis is quite robust towards slow measurement rates. (E): Visitation analysis on simulations using monomers with different degrees of stochastic labeling. The stochastic labeling efficiency (LE) should be slightly below 100% for the correct estimation of the nucleus size. (F): Visitation analysis of simulations considering the influence of photobleaching. The early steps during nucleation can be resolved even at a photobleaching rate of 3% of the association rate. (G): Number of upward and downward transitions starting from different oligomer sizes (monomer numbers).

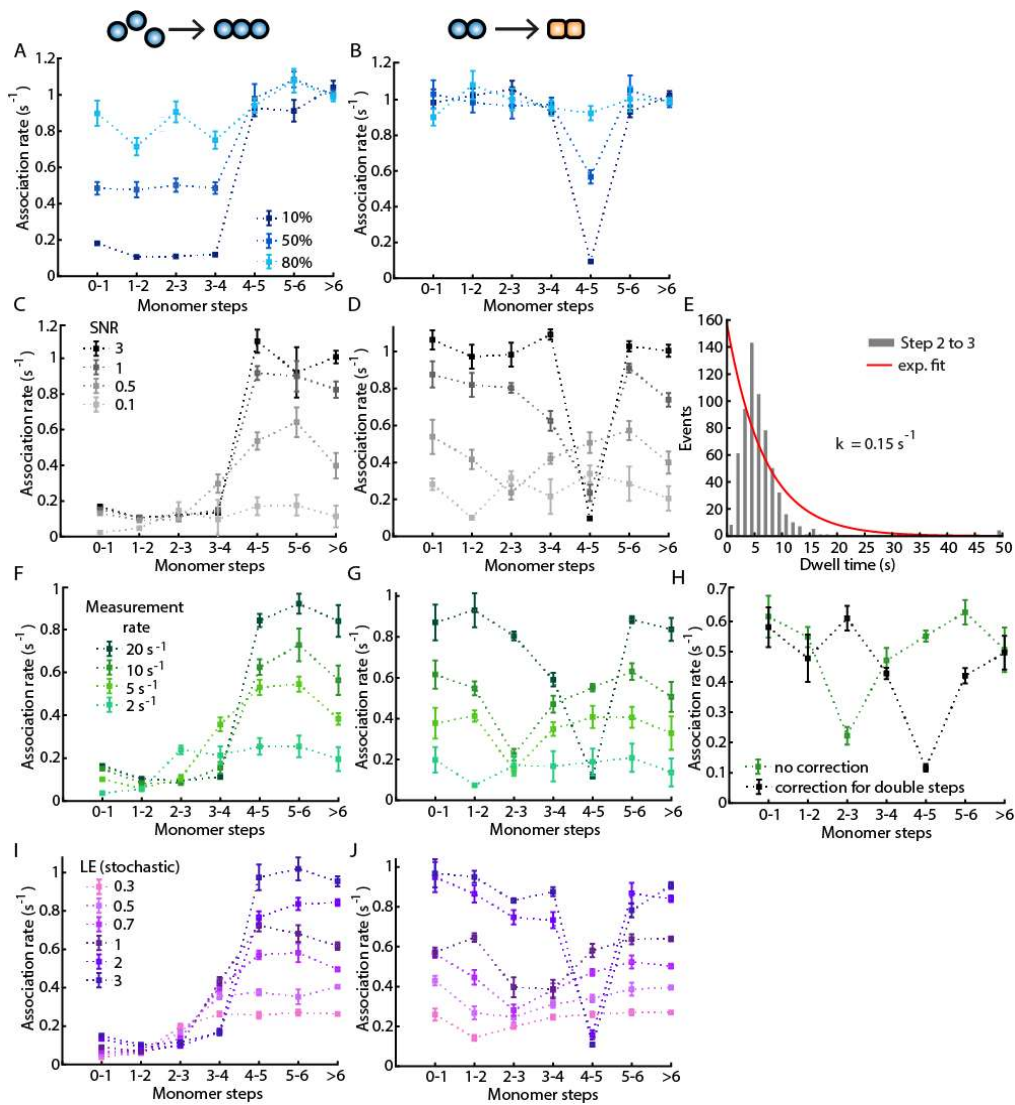

Figure S8: Robustness of the dwell-time analysis. For all simulations, a nucleus or conversion step at four monomers was chosen. (A, B): Extracted on-rates for simulations varying the difference between nucleation and polymerization kinetics. The association rates $k^+_{nuc}$ and $k^+_{conv}$ were chosen to be 10, 50 or 80% of $k^+_{poly}$. (C, D): Extracted on-rates for simulations applying different SNRs. A nucleation mechanism (C) or a slow conversion step (D) can be detected until a SNR of 1. (E): With low SNR, fast steps are not detected and the dwell-time distribution does not follow an exponential decay (red line). (F, G): Extracted on-rates for simulations varying the sampling time. To estimate association rates, the measurement rates should be at least 10x faster than the fastest rate. (H): By using a correction for fast subsequent steps that are detected as single steps with larger amplitude, the correct conversion step can be determined despite a measurement rate of 10 s$^{-1}$ or kinetics rates 10x slower than the sampling rate. (I, J): Extracted on-rates for simulations of monomers with different degrees of labeling. A high labeling efficiency (LE) is needed for correct interpretation of the dwell-time analysis. For stochastic labeling, the LE should ideally be more than 100%.

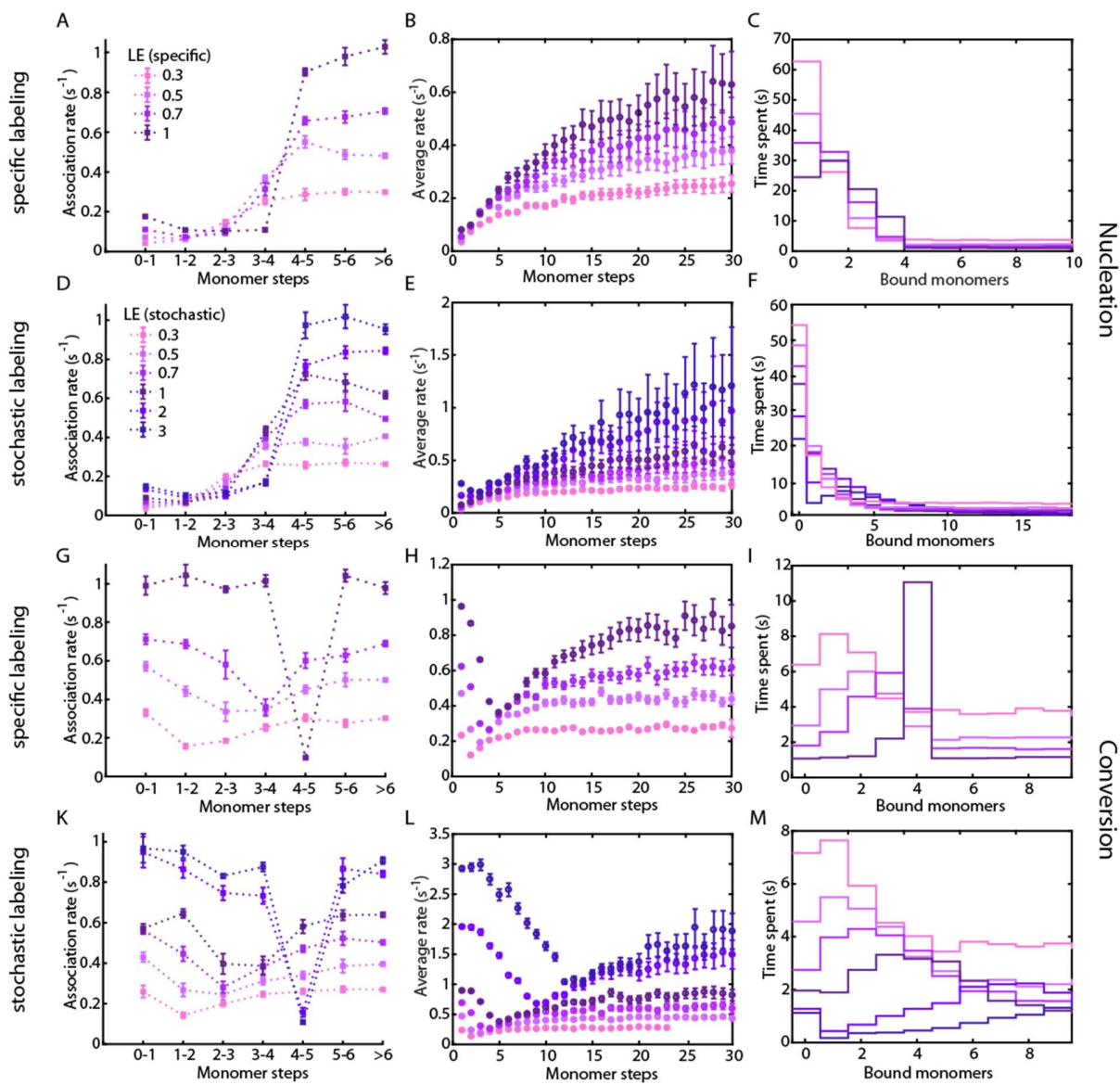

Figure S9: A comparison of the effect of stochastic and specific labeling efficiencies on the different analysis methods. (A-F): A dwell time analysis (A, D), average-rate analysis (B, E) and a visitation analysis (C, F) for simulations of a nucleation mechanism assuming incomplete specific labeling (A-C)) or different degrees of stochastic labeling (D-F). (G-M): A dwell time analysis (G, K), average-rate analysis (H, L) and a visitation analysis (I, M) for simulations of a conversion mechanism assuming incomplete specific labeling (G-I) or different degrees of stochastic labeling (K-M). For specific labeling, there is either 1 or 0 labels per monomer while, for stochastic labeling, multiple labels per monomer are possible. For a nucleation mechanism, a wide range of labeling efficiencies can be tolerated for all analysis methods. However, the dwell-time analysis extracts the correct polymerization rates only for labeling efficiencies of 100% for specific labeling (A) or more than 100% for stochastic labeling (D). For the determination of the correct conversion size, the labeling efficiency should be at 100% for the average rate analysis and the visitation analysis. For the dwell-time analysis, labeling efficiencies at or above 100% are needed.

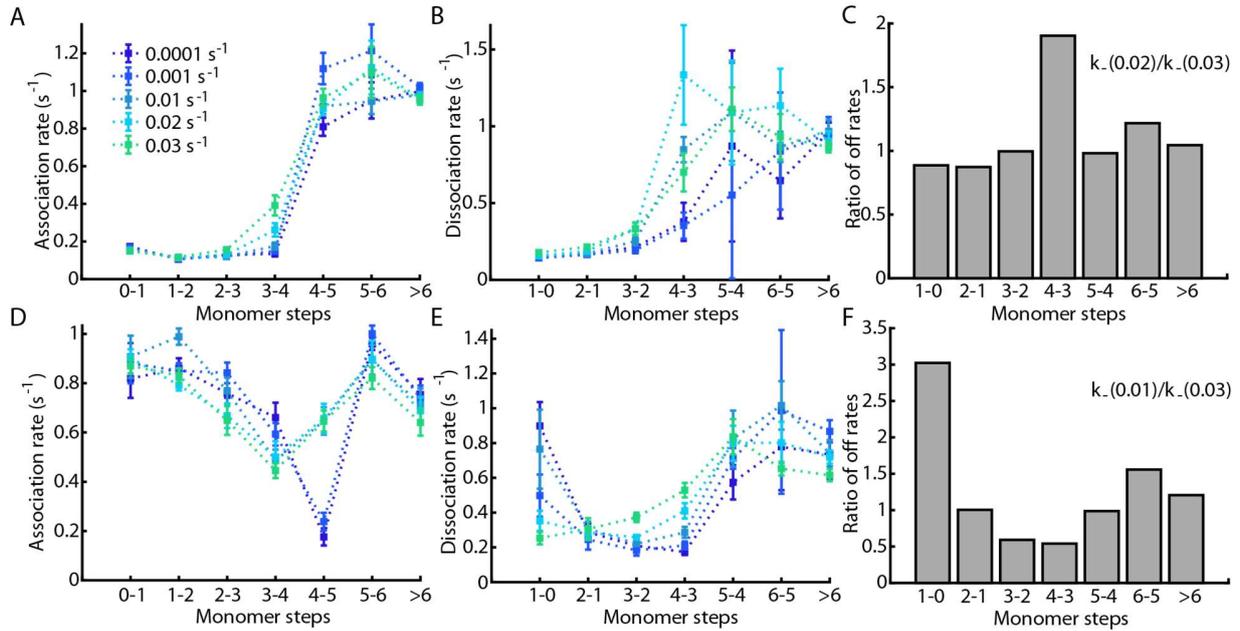

Figure S10: Using photobleaching as a tool to estimate the nucleation or conversion step. Dwell-time analysis on a simulation assuming a nucleation mechanism (A-C) or a conversion mechanism (D-F) with $k^+_{nuc}$ = 0.1 $k^+_{poly}$ and no change in the off-rates. Photobleaching with different photobleaching rates has been applied to the simulations. (A, B): The extracted association rates are plotted and the nucleation or conversion step at 4 monomers is observable. (B, E): The extracted dissociation rates are plotted for the same simulated processes as in (A) and (D). The apparent dissociation rates at the nucleus or conversion size are affected most by different photobleaching rates. (C, F): The ratio between the apparent dissociation rates for the same simulation under the influence of different photobleaching rates are plotted for each step. The difference between the apparent dissociation rates is the highest at the nucleation or conversion size.

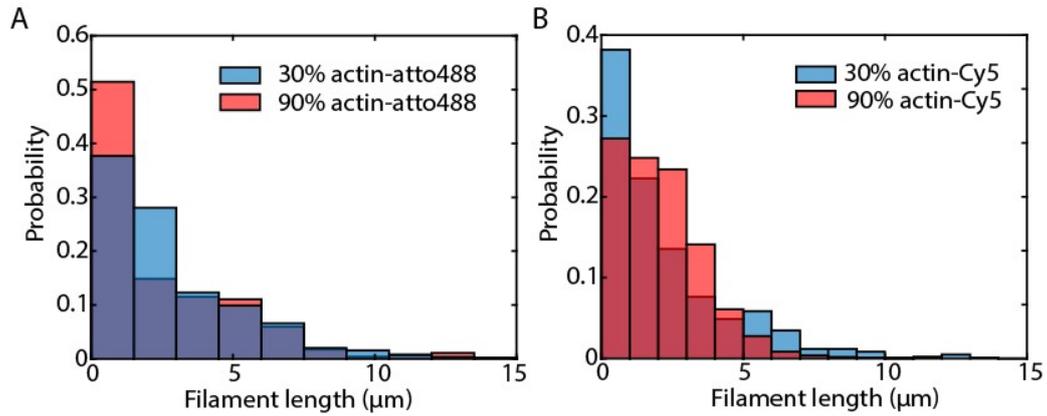

Figure S 11: Actin-Cy5 is functional according to a TIRF assay. A: Filament length distribution 250 s after induction of polymerization of 30% and 90% labeled actin-atto488. Actin-atto488 has been shown to be functional (*6, 37*). B: Filament length distribution of actin-Cy5 at the same time point after induction of polymerization. 30%-labeled actin-Cy5 shows a similar length distribution as actin-atto488.

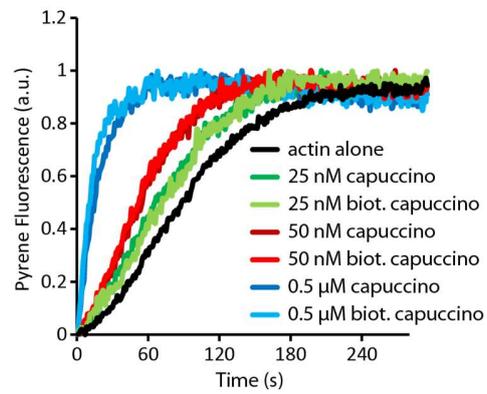

Figure S12: Biotinylated Cappuccino is functional as monitored by a pyrene assay. The nucleation rate of actin alone (black) is enhanced by Cappuccino (dark green, red and blue) and biotinylated Cappuccino (light green, red and blue) in a concentration-dependent manner.

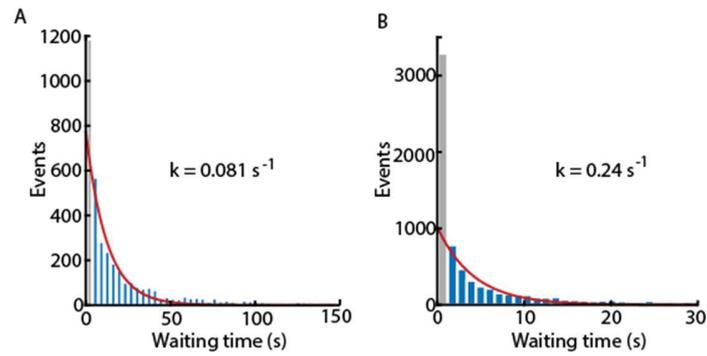

Figure S13: Waiting time distribution of Cappuccino-mediated growth with exponential fit (red) of all steps measured with (A) 5 Hz or (B) 16.7 Hz. The first bin (grey) was not included in the fit, since it shows higher numbers because of the double step correction, where unresolved steps get the interframe time of the measurement as dwell time.

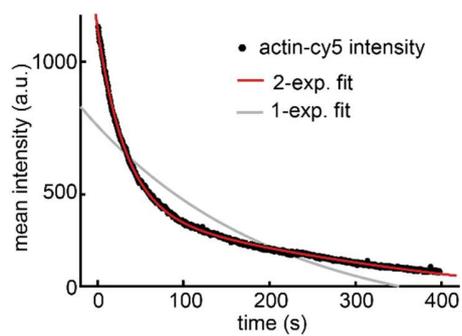

Figure S14: Decay of the average intensity of actin-Cy5 stabilized with phalloidin (see materials and methods for details). A single exponential fit (grey) does not describe the data, but a double-exponential decay (red) does. The decay constants are 0.025 s$^{-1}$ and 0.0015 s$^{-1}$.